\documentstyle[amssymb,prb,aps]{revtex}

\begin{document}
\title{Anelastic relaxation process of polaronic origin in La$_{2-x}$Sr$_{x}$CuO$%
_{4}$: interaction between the charge stripes and pinning centers.}
\author{F. Cordero$^{1}$, A. Paolone$^{2}$, R. Cantelli$^{2}$, M. Ferretti$^{3}$}
\address{$^{1}$ CNR, Area di Ricerca di Tor Vergata, Istituto di Acustica 
``O.M.Corbino``,\\
Via del Fosso del Cavaliere 100, I-00133 Roma, and INFM, Italy}
\address{$^{2}$ Universit\`{a} di Roma ``La Sapienza``, Dipartimento di Fisica,\\
P.le A. Moro 2, I-00185 Roma, and INFM, Italy}
\address{$^{3}$ Universit\`{a} di Genova, Dipartimento di Chimica e Chimica
Industriale,\\
Via Dodecanneso 31, I-16146 Genova, and INFM, Italy}
\maketitle

\begin{abstract}
The evolution of an anelastic relaxation process occurring around 80~K in La$%
_{2-x}$Sr$_{x}$CuO$_{4}$ at a measuring frequency of $\sim 1$~kHz has been
followed from $x=0.0075$ to the overdoped region, $x=0.2$, where it
disappears. The dependence of the peak intensity on doping is consistent
with a polaronic mechanism, identified with the disordered charge stripes
overcoming pinning centers. A marked decrease of the peak amplitude and of the
effective energy barrier for relaxation occurs at $x>0.045$, the same doping
range where a change of the stripe order from parallel to diagonal with
respect to the Cu-O bonds has been observed by neutron diffraction. Both the
energy barrier and peak amplitude also exhibit a rise near $x=1/8$.
\end{abstract}

\twocolumn

\section{INTRODUCTION}

Much evidence has been accumulated of the segregation of the charge carriers
into fluctuating stripes in the cuprate superconductors.\cite
{TSA95,TIU99,WSE99,WBK00,RBC98} In La$_{2-x}$Sr$_{x}$CuO$_{4}$ codoped with
Nd the stripes are seen by neutron diffraction as a static lattice
modulation coherent with the low-temperature tetragonal (LTT)\ modulation;%
\cite{TSA95,TIU99} in the absence of a stable LTT phase, dynamic charge and
magnetic correlations are observed with a spacing incommensurate with the
lattice parameter, and with the direction of the modulation which changes
from diagonal to parallel with respect to the Cu-O bonds when increasing
doping above 6\%.\cite{WBK00} In addition, there are several other
experimental indications that, although not providing a spatial
characterization as the diffraction methods do, still are consistent with
the segregation of the charge carriers into stripes.\cite
{RBC98,BSR96,HSD00,Mar97} In particular, the anelastic spectroscopy has
recently put in evidence a rise of the elastic energy loss in concomitance
with the freezing of the Cu$^{2+}$ spins into the cluster spin glass phase,
which has been interpreted as the motion of the stripes between the pinning
points constituted by the Sr dopants.\cite{79,90} It has also been shown
that the dynamics of such stripes is frozen by the LTT modulation in LBCO
also far from the condition $x=1/8$ for the commensuration between lattice
and stripe spacing.\cite{90} Recently, the issue has been addressed whether
it is possible to observe at higher temperature the process of depinning of
the stripes from the impurities.\cite{82} In fact, among some anelastic
relaxation processes which are observed in La$_{2-x}$Sr$_{x}$CuO$_{4}$
between 50~K and 100~K, there is one at 80~K (measured at 1 kHz) that seemed
to also have the $^{139}$La nuclear quadrupolar relaxation counterpart at
150~K (20~MHz).\cite{82} The latter relaxation process was identified as
magnetic, thereby providing evidence of simultaneous lattice and magnetic
relaxation, unless due to the nearby N\'{e}el transition. Further
measurements have confirmed the second alternative,\cite{unpublished} but
this does not exclude the assignment of the anelastic relaxation to the
stripe depinning. The reason why such a process, together with others, would
be seen by anelastic but not by NQR relaxation is that the latter is
dominated by the relaxation from the Cu spin fluctuations, while the
anelastic spectroscopy is only indirectly sensitive to the spin fluctuations
in case of magnetoelastic coupling.

In what follows a study is presented of the anelastic relaxation process
around 80~K in La$_{2-x}$Sr$_{x}$CuO$_{4}$ over a wide doping range. It is
shown that the process disappears in the overdoped state, as expected from a
polaron-like phenomenon, and its dependence on doping is inconsistent with
explanations in terms of unwanted impurities or off-stoichiometry defects;
it is then discussed in terms of interaction between charge stripes and
pinning centers.

\section{EXPERIMENTAL}

A\ series of ceramic samples of La$_{2-x}$Sr$_{x}$CuO$_{4}$ with $x\le 0.20$
was prepared by conventional solid state reaction as described in Ref. %
\onlinecite{NGM99}. After sintering, the samples were cut in bars
approximately $40\times 4\times 0.6$~mm$^{3}$. Interstitial O is always
present in the as-prepared state of samples with low doping ($x<0.045$), as
evidenced by the anelastic spectra;\cite{79,61} in these cases the excess O
was outgassed by heating in vacuum up to $830~$K.

The complex Young's modulus $E\left( \omega \right) =E^{\prime }+iE^{\prime
\prime }$, was measured as a function of temperature by suspending the bars
on thin thermocouple wires and electrostatically exciting their flexural
modes. The frequencies $\omega _{i}/2\pi $ of the first three odd flexural
modes are in the ratios $1:5.4:13.3$ with $\omega _{1}/2\pi \sim 1$~kHz for
the present samples. The elastic energy loss coefficient is commonly
indicated as the reciprocal of the mechanical quality factor:\cite{NB72} $%
Q^{-1}\left( \omega ,T\right) =$ $E^{\prime \prime }/E^{\prime }$, and was
measured from the decay of the free oscillations or from the width of the
resonance peak. The $Q^{-1}\left( \omega ,T\right) $ function is related to
the imaginary part of the dynamic elastic susceptibility (compliance) and
therefore contains contributions from any process coupled to the measured
strain $\varepsilon $ and having spectral weight at $\omega $. At such low
frequencies only relaxation or diffusion processes are relevant; an
elementary relaxation process consisting of jumps or transitions with rate $%
\tau ^{-1}$ between states which differ in anelastic strain by $\Delta
\varepsilon ^{\text{an}}$ contributes with 
\begin{equation}
Q^{-1}\left( \omega ,T\right) =\Delta \frac{\omega \tau }{1+\left( \omega
\tau \right) ^{2}}\,,  \label{Q-1}
\end{equation}
where the relaxation strength is $\Delta \propto $ $\left( \Delta
\varepsilon ^{\text{an}}\right) ^{2}/T$. The above expression is peaked at $%
\omega \tau =1$, and since the measurements are made at the resonance
frequencies $\omega _{i}$ as a function of temperature, one finds peaks at
the temperatures $T_{i}$ such that $\omega _{i}\tau \left( T_{i}\right) =1$.
Therefore, measurements at different frequencies provide the temperature
dependence of the rate $\tau ^{-1}\left( T\right) $, which generally follows
the Arrhenius law 
\begin{equation}
\tau ^{-1}\left( T\right) =\tau _{0}^{-1}\exp \left( -E/k_{\text{B}}T\right)
\,.  \label{Arrh}
\end{equation}
A process characterized by a spectrum of relaxation times may be described
by the phenomenological expression 
\begin{equation}
Q^{-1}\left( \omega ,T\right) =\Delta \frac{1}{\left( \omega \tau \right)
^{\alpha }+\left( \omega \tau \right) ^{-\beta }}\,,  \label{Q-1ab}
\end{equation}
which is the Fuoss-Kirkwood expression for $\beta =\alpha <1$,\cite{NB72}
and reduces to the Debye formula, Eq. (\ref{Q-1}), for $\alpha =\beta =1$.
The parameter $\alpha $ controls the peak broadening in the low temperature
region, where $\omega \tau <1$, and $\beta $ in the high temperature region.

\section{RESULTS}

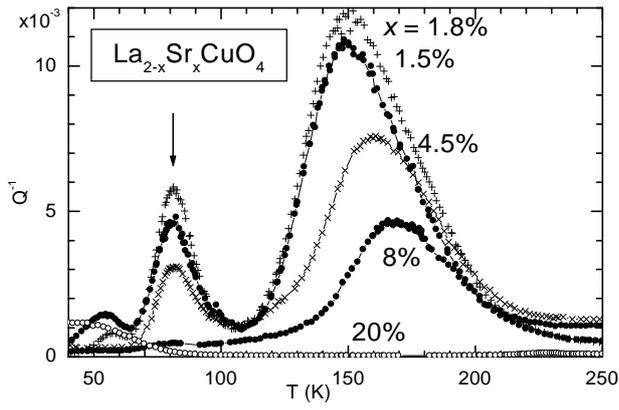
\begin{figure}[tbp]
\caption{Elastic energy loss coefficient versus temperature of La$_{2-x}$Sr$%
_{x}$CuO$_{4}$ with $x=0.015$ (1.3~kHz), $x=0.018$ (1.0~kHz), $x=0.045$
(0.85~kHz), $x=0.08$ (1.0~kHz) and $x=0.20$ (2.7~kHz). The peak attributed
to the interaction between stripes and pinning centers is indicated with an
arrow.}
\label{spectra}
\end{figure}

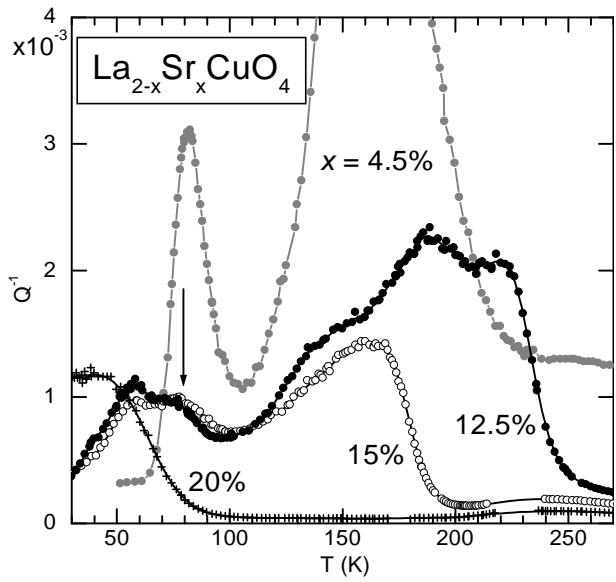
\begin{figure}[tbp]
\caption{Elastic energy loss coefficient versus temperature of La$_{2-x}$Sr$%
_{x}$CuO$_{4}$ with $x=0.045$ (0.85~kHz), $x=0.125$ (1.5~kHz), $x=0.15$
(1.7~kHz) and $x=0.20$ (2.7~kHz). The peak attributed to the interaction
between stripes and pinning centers is indicated with an arrow.}
\label{opt/over}
\end{figure}

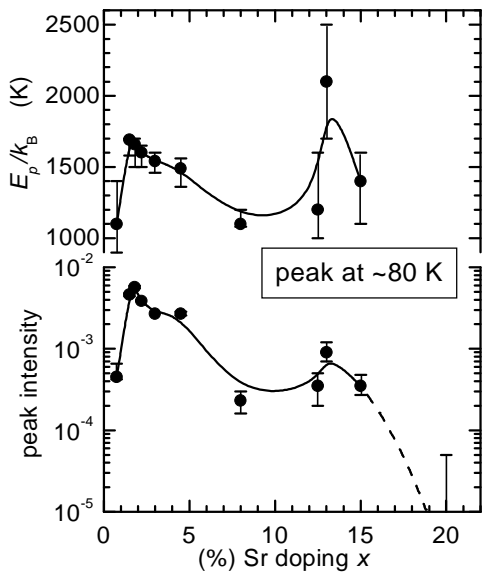
\begin{figure}[]
\caption{Lower panel: intensity of the elastic energy loss peak at $\sim 80$
~K as a function of Sr doping. Upper panel: apparent activation energy of
the peak. The continuous lines are guides for the eye.}
\label{Qmvsx}
\end{figure}

Figure \ref{spectra} presents the anelastic spectra of La$_{2-x}$Sr$_{x}$CuO$%
_{4}$ at various dopings, measured exciting the first flexural modes at $%
\sim 1$~kHz. The peak around 150~K becomes much more intense in La$_{2}$CuO$%
_{4}$ free of interstitial O,\cite{61} where it can also be observed as a
maximum in the $^{139}$La nuclear quadrupolar relaxation rate.\cite{68} It
has been associated with the collective tilting dynamics of the O octahedra
in local multiwell potentials, describable as soliton-like tilt waves.\cite
{68} The depression of the peak intensity and perhaps also the shift to
higher temperature with increasing doping are connected with the lattice
disorder introduced by\ Sr, which hinders the collective tilt modes; this
process will not be considered further here. Another minor peak may appear
near 50~K, whose behavior as a function of Sr and O doping is less regular
than that of the other two processes, and this peak also will not be
considered further. The interest will be focused on the peak near 80~K,
indicated with an arrow; its intensity becomes rather small for $x\ge 0.08$
and completely disappears in the overdoped state $x=0.20$. The spectra at
optimal and higher doping are shown separately in Fig. \ref{opt/over}; the
step above 220, 170 and 50~K in the $Q^{-1}\left( T\right) $ curves at $%
x=12.5$, 15 and 0.20 is due to the tetragonal/orthorhombic (HTT/LTO)
structural transformation,\cite{61} which shifts to lower temperature with
increasing doping. The curve for $x=0.045$ has also been included, in order
to render the peak near 80~K more recognizable (indicated by an arrow). The
dependence of the intensity of the 80~K maximum on doping, $Q_{M}^{-1}\left(
x\right) $, is shown in Fig. \ref{Qmvsx} in logarithmic scale, together with
other concentrations not reported in Figs. \ref{spectra} and \ref{opt/over}.

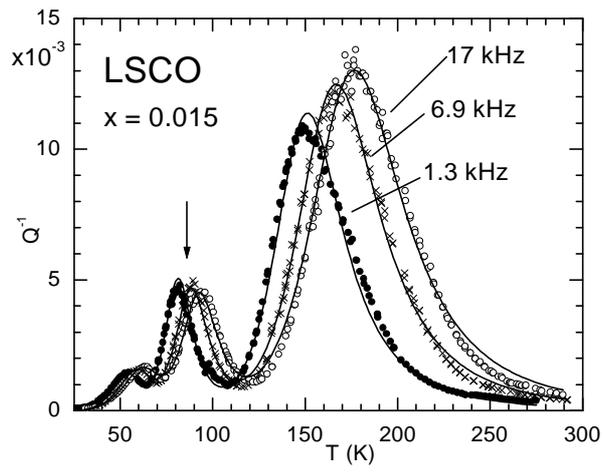
\begin{figure}[]
\caption{Anelastic spectrum of La$_{2-x}$Sr$_{x}$CuO$_{4}$ with $x=0.015$
measured on the 1st, 2nd and 3rd flexural modes at the frequencies indicated
in the figure. The continuous lines are fits as described in the text. The
peak attributed to the interaction between stripes and pinning centers is
indicated with an arrow.}
\label{LSCO15}
\end{figure}

Figure \ref{LSCO15} shows a fit of the spectra of the $x=0.015$ sample
measured at three frequencies. All the peaks are fitted with the expressions
(\ref{Q-1ab}) and (\ref{Arrh}); for the 50~K and 150~K peaks, an additional
temperature dependence of the relaxation strength had to be introduced. The
analysis of the 80~K peak, like in Fig. \ref{LSCO15}, can be properly
carried out up to $x=0.045$, while at higher doping the peak becomes rather
small and masked by the HTT/LTO structural transition and the other peak
near 50~K, so that only its height can be extracted without an appreciable
error. The peak is rather broad, with a width parameter of the
Fuoss-Kirkwood distribution $\alpha =\beta =0.6-0.7$. The upper panel of
Fig. \ref{Qmvsx} shows the dependence of the activation energy $E_{{\rm p}}$
on doping; the most reliable data, based on analyses like that in Fig. \ref
{LSCO15}, show a decrease from $E_{{\rm p}}/k_{\text{B}}=$ 1700~K to 1100~K
when $x$ passes from 0.015 to 0.08, while the preexponential factor $\tau
_{0}$ passes from $1.4\times 10^{-13}$~s to $10^{-10}$~s, so that the
temperature of the peak measured around 1~kHz does not change much; for this
reason we will refer to the ''80~K peak'', but it is understood that the
peak temperature is different at different frequencies. The data at $x=0.0075
$ and $x>0.08$ have larger error bars, but sufficient to reveal a remarkable
resemblance of the $E_{{\rm p}}\left( x\right) $ curve with $\log \left[
Q_{M}^{-1}\left( x\right) \right] $ in the lower panel,\cite{note}
including a peak around $x=1/8$.

\section{DISCUSSION}

Before discussing the 80~K relaxation process in terms of stripes, the
possibility should be considered that it is due to some impurities or off
stoichiometry defects. The explanation in terms of foreign impurities is
very unlikely in view of the dependence of the peak intensity $%
Q_{M}^{-1}\left( x\right) $ on doping (see Fig. \ref{Qmvsx}). If the cause
were an impurity in the starting CuO powder, like Fe, the effect should be
independent of doping, or possibly depend in a monotonous way on $x$;
neither can a correlation be found between the $Q_{M}^{-1}\left( x\right) $
and $x$ or $1-x$, which might justify an impurity in the starting La or Sr
oxides. The O off-stoichiometry has also to be considered, since it is
certainly present. Interstitial O\ cannot be the cause of the 80~K peak,
which is even suppressed at high enough O content (oxygenated or as-prepared
state at low $x$, not shown here). Regarding the O vacancies in the CuO$_{2}$
planes, they may certainly be present in some of the outgassed samples\cite
{67} with $x\le 0.045$, but the other samples have not been outgassed
because no traces of interstitial O were found in the as-prepared state.
Therefore we conclude that no impurity or defect connected with
off-stoichiometry can account for the $Q_{M}^{-1}\left( x\right) $
dependence in Fig. \ref{Qmvsx}.

Instead, the dependence of the intensity of the 80~K peak on doping is fully
consistent with a polaronic origin. In fact, the process should be absent in
the undoped state, and indeed a steep decrease of the intensity occurs below
1.8\% doping. It should also disappear in the overdoped state, where the
charge carriers are in a uniform metallic state, and in fact the peak is
undetectable at $x=0.20$, after passing through a maximum around $x\sim 0.02$
and a secondary maximum near $x=1/8$. A truly undoped state is
difficult to obtain, due to the presence of both interstitial O and O
vacancies. The first is readily detected by the intense $Q^{-1}\left(
T\right) $ peaks due to its hopping;\cite{61,63} The typical content of
interstitial O at $x=0$ after sintering is\cite{67} $\lesssim 0.005$\ and
causes a doping of $\lesssim 0.01$; as mentioned above, the samples with $%
x\le 0.045$ were subjected to an outgassing treatment, but at $x=0$ it is
difficult to completely eliminate interstitial O without also creating O
vacancies in the CuO$_{2}$\ planes.\cite{67}. To our knowledge, the effect
of these vacancies on the electric and magnetic properties has not been
studied, but the anelastic spectra depend critically on the presence of such
O defects in the $50-250$~K range. For this reason, we could not obtain a
sample with $x=0$ which is at the same time completely undoped and without a
peak around 80~K.

Regarding the nature of the process, if the peak were only observed at very
low doping, it might be associated with a localized defect, like a small
polaron hopping around a particular type of impurity, but the persistence of
the peak at doping levels as high as 0.15 excludes this possibility. At high
doping the only reasonable possibility of a polaronic relaxation is
connected with the charge stripes, and, since thermal activation over a
barrier of $E_{{\rm p}}\ge 0.1$~eV is required, the process must be
connected to the presence of pinning centers. In fact, although the dynamics
of the stripes has never been measured precisely, their motion far from
pinning centers must be much faster than the average $\tau ^{-1}$ of the
80~K relaxation process: the anelastic spectra at liquid He temperatures\cite
{79,90} indicate that the mean fluctuation frequency passes through the kHz
range close to the $T_{g}\left( x\right) \simeq $ $0.2~$K$/x$ temperature
for the freezing into the cluster spin glass state. Probes with higher
characteristic frequencies, like NQR and $\mu $SR detect onset temperatures
for the stripe ordering with progressively higher temperatures, up to $\sim
20$~K for the neutron scattering experiments,\cite{WSE99} and the effective
energy barriers for such fluctuations are in the range $E=20-60$~K, as
discussed in Ref. \onlinecite{92}.

Therefore, the proposed picture is that the charge stripes may fluctuate
very fast far from the pinning centers, as it appears from most experiments,%
\cite{WSE99,RBC98,79,90,92} but the process of depinning or overcoming the
pinning centers requires the energy $E_{{\rm p}}\ge 0.1$~eV deduced from
the 80~K peak. The regularly spaced arrays of stripes are observed by
diffraction well below 80~K,\cite{TSA95,TIU99,WSE99,WBK00} but around this
temperature the stripes would already exist in a disordered state, which can
be identified with the nematic stripe phase devised by Kivelson {\it et al.}%
\cite{KFE98}.

The task of discussing the doping dependence of the relaxation strength of
this depinning process in terms of some existing model, like those of a
moving dislocation or domain wall, is not obvious. In fact, such models are
appropriate when interpreting the anelastic relaxation at liquid He
temperatures in the cluster spin glass phase,\cite{79,90} where the stripes
act as walls between domains with AF\ correlations, and the stripe motion is
associated with a variation in the sizes of the adjacent domains. If an
anisotropic strain is associated with the staggered magnetization within
each domain, than the change of the sizes of domains with different
orientations causes a change of the anelastic strain $\varepsilon ^{\text{an}%
}$. Similarly, the movement of a dislocation is associated with the shift of
a part of the crystal, and again is a source of anelastic strain. Instead, a
charge stripe at 80~K does not separate domains with different anelastic
strains, since at these temperatures the Cu spins are fluctuating much
faster than the sample vibration frequency;\cite{RBC98} therefore, it is no
more appropriate to associate an anelastic strain proportional to the area
swept by the stripe, as in the above models. In the present case, $%
\varepsilon ^{\text{an}}$ should be associated with the different
configuration of the stripe in the pinned or unpinned state, or on either
side of the pinning center. It is likely that $\varepsilon ^{\text{an}}$,
like the distortion coupled to the presence of holes,\cite{Mar97} is mainly
connected with the in-plane shear also associated with the tilts of the
octahedra. Then, it is reasonable to assume that $\varepsilon ^{\text{an}}$
is an increasing function of the local degree of tilting of the octahedra.
This would partly explain the general reduction of the relaxation strength
with increasing doping, because both the in-plane shear strain and tilt
angles $\Phi $ of the octahedra are decreasing functions of doping.\cite
{BBK99} Note, however, that also other ingredients must determine the
relaxation strength, since the marked decrease of $Q_{M}^{-1}\left( x\right) 
$ between $x=0.05$ and 0.08, would not be justified by the smooth $\Phi
\left( x\right) $ function.\cite{BBK99} Interestingly, the jump of $%
Q_{M}^{-1}\left( x\right) $ occurs in the doping region of the crossover
from parallel to diagonal stripes,\cite{WBK00} providing an additional
indication that the 80~K peak is indeed associated with the stripes.

Regarding the relaxation rates $\tau ^{-1}\left( T\right) $ deduced from
fits like that in Fig. \ref{LSCO15}, the values of the preexponential
factors, $\tau _{0}^{-1}=$ $10^{10}-10^{13}$~s$^{-1}$, may seem too small
for a polaronic relaxation, but are reasonable for an extended entity like
the charge stripe. The most striking observation is the close similarity
between the $E_{{\rm p}}\left( x\right) $ and the $\log \left[
Q_{M}^{-1}\left( x\right) \right] $ curves, with a sharp peak centred at $%
x=0.13$, near $x=1/8$, out of a general decrease with increasing doping. The
activation energy $E_{{\rm p}}$ can be identified with the barrier for
depinning or overcoming the pinning center and must therefore depend on the
nature of this defect. Two possibilities are the Sr dopants or the twin
boundaries between the two variants of the LTO domains; the latter are
immobile at 80~K, since the motions of the twin walls and of the tilt waves
are observed in the anelastic spectra just below the HTT/LTO transition and
at 150~K respectively.\cite{61}

The hypothesis that pinning is due to irregularities in the energy landscape
resulting from the octahedral tilts, e.g. the twin walls and/or the
substitutional disorder in the La/Sr sublattice, provides an
explanation for $E_{{\rm p}}\left( x\right) $ being a generally decreasing
function of $x$. In fact, the decrease of the tilt angles $\Phi \left(
x\right) $ implies closer minima in the multiwell potential for the
octahedral tilts, smaller barriers and therefore a smoother overall
potential. The rather broad distribution of relaxation rates ($\alpha
=0.6-0.7$ instead of 1 in Eq. \ref{Q-1ab}) is also understandable in this
picture, since the pinning energy would depend on the particular geometry of
the local tilts. The close relationship between the $E_{{\rm p}}\left(
x\right) $ and the $\log \left[ Q_{M}^{-1}\left( x\right) \right] $ curves
can also be understood in this framework, since stripes that are more
defined and associated with larger atomic displacements must result both in
larger relaxation strength and larger pinning energy; in particular, the
rise of both near $x=1/8$ is consistent with the tendency of the lattice to
develop an LTT modulation commensurate with the stripes,\cite{TSA95,TIU99}
which makes them more well defined but also tends to pin them.

\section{CONCLUSION}

The anelastic relaxation process occurring in La$_{2-x}$Sr$_{x}$CuO$_{4}$
around 80~K for $\omega /2\pi \sim 1$~kHz has been studied over a wide
doping range. All the features of this relaxation process, and notably the
dependence of its intensity and activation energy on doping, confirm the
earlier proposal that it is associated with disordered charge stripes
interacting with pinning centers. Within this interpretation, the activation
energy measured from the dependence of the anelastic spectra on frequency, $%
E_{{\rm p}}\sim 0.1-0.18$~eV, represents the energy necessary for the
stripes to overcome the pinning centers, and is enhanced in correspondence
with the doping $x\sim 1/8$, at which the stripe and lattice modulations become
commensurate. The pinning potential can be associated with irregularities in
the ordered pattern of the octahedral tilts due to the Sr dopants and/or
other defects that are frozen below 100~K, like the twin walls.

\section*{Acknowledgments}

The authors thank Prof. A. Rigamonti for useful discussions.


\begin{references}
\bibitem{note}  The error bars refer to the analysis of one sample
(sometimes two samples) nearly free of interstitial O at each doping; the
intensity of the peak depends on the exact O stoichiometry, especially at
low doping, so that larger error bars might result from an extensive
analysis of several samples.
\end{references}

\end{document}